\documentclass[10pt,conference]{IEEEtran}\IEEEoverridecommandlockouts

\usepackage{amsfonts}
\usepackage{slashbox}
\usepackage{amsmath}
\usepackage{multirow}
\usepackage{graphicx}
\usepackage{amsfonts}
\usepackage{amsmath,epsfig}
\usepackage{mathbbold}
\usepackage{stfloats}
\usepackage{amssymb}
\usepackage{url}
\usepackage{bm}
\usepackage{cite}
\usepackage{amsmath}
\usepackage{amssymb}
\usepackage{latexsym}
\usepackage{multirow}
\usepackage{epsfig}
\usepackage{graphics}
\usepackage{mathrsfs}
\usepackage{algorithmic}
\usepackage{algorithm}
\usepackage{subfigure}
\usepackage{slashbox}
\usepackage[all]{xy}

\usepackage{pifont}
\usepackage{bbding}
\usepackage{stmaryrd}
\usepackage{amssymb}
\usepackage{amsfonts}
\usepackage{epic}
\usepackage{stfloats}
\usepackage{latexsym}
\usepackage{epstopdf}
\usepackage{epic}
\usepackage{bm}
\usepackage{xcolor}
\usepackage{mathrsfs}
\usepackage{pifont}
\usepackage{bbding}
\usepackage{amsmath,epsfig}
\usepackage{mathbbold}
\usepackage{stmaryrd}
\usepackage{amssymb}
\usepackage{amsfonts}
\usepackage{epic}
\usepackage{graphicx}
\usepackage{subfigure}

\usepackage{enumerate}

\usepackage{stfloats}
\usepackage{latexsym}
\usepackage{epstopdf}
\usepackage{epic}
\usepackage{multirow}
\usepackage{stfloats}
\usepackage{bm}

\usepackage{booktabs}
\usepackage{color}

\newcommand{\q}{\mathbf q}
\newcommand{\Q}{\mathbf Q}
\newcommand{\A}{\mathbf A}
\newcommand{\w}{\mathbf w}

\begin{document}

\title{Joint Trajectory and Communication Design for UAV-Enabled Multiple Access}
% Overlaying cooperative cellular networks:  Energy-Efficient D2D Communications with spectrum-power trading
%\author{\IEEEauthorblockN{Qingqing Wu, \emph{Student Member, IEEE},  Geoffrey Ye Li, \emph{Fellow, IEEE},  \IEEEauthorblockN{Wen Chen},  \emph{Senior Member, IEEE}, and \IEEEauthorblockN{Derrick Wing Kwan Ng},  \emph{Member, IEEE}%, \emph{Member, IEEE},
%\thanks{ Qingqing Wu and Geoffrey Ye Li are with the School of Electrical and Computer Engineering, Georgia Institute of Technology, USA, email:\{qingqing.wu, liye\}@ece.gatech.edu. Wen Chen is with Department of Electronic Engineering, Shanghai Jiao Tong University, China, email: \{wenchen\}@sjtu.edu.cn. Derrick Wing Kwan Ng is with the School of Electrical Engineering and Telecommunications, The University of New South Wales, Australia, email: w.k.ng@unsw.edu.au. }}  }

\author{\IEEEauthorblockN{Qingqing Wu, Yong Zeng, and Rui Zhang}
\IEEEauthorblockA{ Department of Electrical and Computer Engineering, National University of Singapore}
Emails: \{elewuqq, elezeng, elezhang\}@nus.edu.sg}

\maketitle
%\begin{abstract}
%\end{abstract}
%
%\begin{keywords}
%\end{keywords}
%%\newpage
\vspace{-0.2cm}
\begin{abstract}
Unmanned aerial vehicles (UAVs) have attracted significant interest recently in wireless communication due to their high maneuverability, flexible deployment, and low cost. This paper studies a UAV-enabled wireless  network where the UAV is employed as an aerial mobile  base station (BS) to serve a group of users on the ground. To achieve fair performance among users, we maximize  the minimum  throughput over all ground users by jointly optimizing the multiuser communication scheduling and UAV trajectory over a finite horizon. The formulated  problem is shown to be a mixed integer non-convex optimization problem that is difficult to solve in general.  We thus  propose an efficient iterative algorithm by applying the block coordinate descent and successive convex optimization techniques, which is guaranteed to converge to at least a locally optimal solution.
%Specifically, the original problem is divided into into two sub-problems regarding to each optimization block, and for each of them, either a relaxed problem or an approximated convex optimization problem is solved by keeping the other block of variables fixed.
%We further show that  the proposed algorithm
To achieve fast convergence and stable throughput, we further propose a low-complexity  initialization scheme for the UAV trajectory design based on  the simple circular trajectory. Extensive simulation results are provided which show significant throughput gains of the proposed design as compared to other benchmark schemes.
%The proposed trajectory design also outperforms the circular trajectory significantly.

%We propose an iterative optimization framework which is guaranteed to converge to a local optimal solution. Specifically,  the original optimization problem is divided into into three sub-problems and in each iteration, we alternatively optimize one blocks of variables while keeping the other two fixed. By exploiting the relaxation and the successive convex optimization technique, all the three sub-problems are transformed into convex optimization problem.
\end{abstract}
\vspace{-0.2cm}
\section{Introduction}
Unmanned aerial vehicles (UAVs) have attracted significant attention in recent years for military as well as various  civilian applications, such as surveillance and monitoring, aerial imaging, cargo delivery, etc.
%This is attributed to the significant advancements in the UAV technology, such as increased payload capacity, longer average flight duration.
 As reported in \cite{globalUAVmarket}, the global market for commercial UAV applications, estimated at about 2 billion US dollars in 2016, will skyrocket to as much as 127 billion US dollars by 2020.
Equipped with advanced transceivers and smart sensors, UAVs are gaining increasing popularity in the information technology (IT) community due to their high maneuverability and flexibility for on-demand deployment. In particular, UAVs typically have high possibility of line-of-sight (LoS) air-to-ground communication links, which is appealing to  the wireless service providers.  Several leading IT companies have launched pilot projects,  such as project Aquila by Facebook \cite{fb_UAV} and Loon by Google \cite{gl_UAV}, for providing ubiquitous internet access worldwide by leveraging the UAV/drone technology. Meanwhile,  extensive research efforts from the academia have also been devoted to employing UAVs as different types of %a wide range of %applications in wireless networks \cite{zeng2016wireless}.
%and are also envisioned as a promising technology for wireless communications due to their capabilities for coverage extension and capacity enhancement, which is particularly important for
% As most aerial work requires low altitude and low speed operation, rotary  wing UAVs are more popular in the field of civilian UAVs.
%Small-scale fixed-wing UAVs as research platforms are generally less popular than the rotorcraft counterparts.
% have found a wide range of applications for wireless networks
%Recently, UAVs have been investigated as different
 %types of
  wireless communication platforms  \cite{zeng2016wireless}, such as aerial mobile base stations (BSs) \cite{mozaffari2015drone,mozaffari2016efficient,al2014optimal,lyu2016placement,bor2016efficient}, mobile relays \cite{zhan2011wireless,zeng2016throughput}, and flying computing cloudlets \cite{jeong2016mobile}. In particular, employing UAVs as aerial BSs is envisioned as a promising solution to enhance the performance  of the existing  cellular systems.
%unmanned aerial vehicles (UAVs) can serve a multitude of purposes such as surveillance,
%localization and communication, making them a flexible solution to augment and enhance the
%capabilities of the current cellular systems
%Employing UAV as a flying base station has many advantages. It is more likely to have a higher possibility of enjoying line-of-sight channels.
%to facilitate temporary hot spots and compensate network outages in case of public events and emergencies
%Use of single or multiple UAVs as communication relays,  aerial base stations,  for network provisioning in emergency situations and for public
%safety communications has been of particular interest due to their fast deployment and large coverage capabilities.
%In general, there are two main types of UAVs, rotary-wing and fixed-wing UAVs. rotary-wing UAV has a better maneuverability.
%UAVs have been exploited to achieve different functionalities for wireless communication systems, such as an aerial BS, a mobile relay, and a flying computing center.
Depending on whether the UAV mobility is exploited or not, two different lines of research can be identified along this direction, i.e., static-UAV or  mobile-UAV enabled
wireless networks.

 The research on the static-UAV enabled  networks mainly focuses on the UAV deployment/placement optimization \cite{al2014optimal,lyu2016placement,bor2016efficient}, with the UAVs serving as aerial quasi-static BSs to support ground users in a given area.
%However, the channel characteristic of the air-to-ground link is quite different from that of the conventional ground-to-ground link. In fact, a higher altitude, although leading to a larger pathloss, also provides a higher probability of enjoying line-of-sight (LoS) channels.
As such, the altitude and the horizonal location of the UAV can be either separately or jointly optimized. The authors in \cite{al2014optimal} provide an analytical approach to optimize the altitude of a UAV for providing maximum coverage for ground users. In contrast, by fixing the altitude, the horizonal positions of UAVs are optimized  in \cite{lyu2016placement} to minimize the number of UAV BSs required to cover a given set of ground users.  A similar problem is also studied in \cite{bor2016efficient} for a drone-enabled small cell placement optimization in three-dimensional (3D)  space.
% The resulting problem is then solved via a numerical approach. by an iterative spiral algorithm
 %By leveraging stochastic geometry, the authors in \cite{chetlur2017downlink} develop a general framework for the downlink coverage analysis of a network with multiple static UAVs.% which also serves as a foundation for the investigation of co-existence of UAV networks and terrestrial cellular networks.
%the donwlink outage probabilities for  modeling the network with multiple UAVs  as a uniform binomial point process,

In addition to the UAV placement optimization,  exploiting the UAV high-mobility in the mobile-UAV enabled networks is anticipated to unlock the full potential of UAV-enabled communications. With the fully controllable UAV mobility,  the communication distance between the UAV and ground users can be significantly shortened by proper UAV trajectory design and communication scheduling. This is analogous and yet in sharp contrast to the existing small-cell technology, where the cell radius is  reduced by increasing the number of small-cell BSs deployed,  but at the cost of increased infrastructure expenditure. Motivated by this, the UAV trajectory optimization problem is rigorously studied in \cite{zeng2016throughput} and \cite{zeng2016energy} for a mobile relaying system and point-to-point energy-efficient system, respectively. To reap the full benefit of UAV mobility, a novel cyclical multiple access scheme is proposed in \cite{lyu2016cyclical}
%where the UAV communicates with each ground user when it flies sufficiently  close to in it a periodic (cyclical) time-division manner.
where an interesting throughput-access delay trade-off is  revealed. Specifically, it has been shown that significant throughput gains can be achieved over the case of a static UAV for delay-tolerant applications. However, in \cite{lyu2016cyclical} the users are assumed to be uniformly located in a one-dimensional (1D) line and the UAV is restricted to fly at a constant speed, which simplifies the analysis but limits the applicability in practice.
%In practice, the user access delay arising from cyclical multiple access can be reduced by various approaches such as increasing the number and/or the maximum speed of UAVs.
%The trajectory design is also studied in \cite{mozaffari2016unmanned} where a UAV is despatched to serve a set of users. However, the UAV is only assumed to transmit signals at some pre-defined locations and the trajectory is thereby heavily restricted while regardless of the transmission, leading to an underestimated performance gain of trajectory design.
%each user is only allowed to communicate with the UAV when the UAV gets close to it.

%Many ways can be employed to reduce the user access delay, such as . It is more suitable to periodic and  delay tolerant services.
%xxx
% \cite{bor2016efficient,gupta2016survey,zhan2011wireless}
% \cite{jeong2016mobile,kandeepan2014aerial,loke2015internet,chen2016caching}
%users are scheduled to communicate with the UAV in a cyclical time-division manner
%However, it has to be note that the performance does not come for free, which is in fact at the increase of user access delay, since each user has to wait for UAV to get close to it.

In this paper, we consider a single UAV-enabled wireless network where the UAV is employed to serve a  group of users in a given two-dimensional (2D) area.  Our goal is to maximize the minimum average rate among all users by jointly optimizing  the user communication scheduling and UAV trajectory in a finite period. Different from \cite{lyu2016cyclical}, we study a general and practical setup where users are freely located on the ground and the UAV trajectory can be optimized in 2D along with the multiuser communication scheduling. Such a joint optimization problem is new and not yet investigated in the literature, to our best knowledge. On one hand, with any given user scheduling,  it is intuitive that the UAV should visit users according to the order that users are scheduled for communication to achieve short-distance links. On the other hand, for any fixed UAV trajectory, the UAV should accordingly  schedule the users for communication based on their distances to it. As a consequence, the user scheduling and UAV trajectory optimization are closely coupled with each other in our considered problem, which makes it challenging to solve optimally in general. To tackle this problem, we first relax the binary variables for user scheduling into continuous variables and solve the resulting problem with  an efficient iterative algorithm devised by leveraging the block coordinate descent method \cite{hong2016unified}.  Specifically,  one of the two blocks of variables for the user scheduling and UAV trajectory is optimized alternately in each iteration, while keeping the other block fixed. However, even for fixed user scheduling, the UAV trajectory optimization problem is still difficult to solve due to its non-convexity. We thus apply the successive convex optimization technique \cite{hong2016unified} to solve it approximately. Our proposed algorithm is guaranteed to converge to at least a locally optimal solution of the joint user scheduling and UAV trajectory design problem. It is shown by simulation  that significant throughput gains are achieved by our proposed joint design, as compared to conventional static UAV or heuristic UAV trajectory benchmarks. It is also observed that the throughput of the proposed mobile UAV system increases with the UAV trajectory period, $T$, showing a peculiar throughput-access delay trade-off \cite{lyu2016cyclical} in UAV-enabled 2D communication.
%As long as the period is sufficiently large, the UAV is able to capture the shortest air-ground communication link, i.e., stay on top of each user for information transmission. %In addition, the throughput-delay tradeoff is demonstrated in the general setup.
\vspace{-0.1cm}
\section{System Model and Problem Formulation}
   \begin{figure}[!t]
\centering
\includegraphics[width=0.25\textwidth]{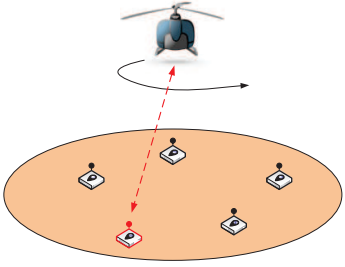}
\caption{ A UAV-enabled wireless network.} \label{UAV}\vspace{-0.4cm}
\end{figure}
\subsection{System Model}
As shown in Fig. 1, we consider a wireless communication system where a UAV is employed as an aerial BS to serve a group of $K$ users on the ground.  The user set is denoted by $\mathcal{K}$ with $|\mathcal{K}|=K$. We study the downlink communication scenario from the UAV to ground users while the obtained results are directly applicable to the uplink transmission from ground users to the UAV as well.  The considered setup could practically correspond to an information dissemination or a data collection system enabled by the UAV.
 Assume that the UAV serves the ground users via a periodic/cyclical time-division multiple access (TDMA) with each period/cycle of duration denoted by  $T$. %Specifically, ....
 Note that the choice of $T$ has a significant impact on the system performance. On one hand, thanks to the UAV mobility, a larger period  $T$ provides more time for the UAV to move closer to each user to achieve better communication channels and hence higher throughput.
 Intuitively, as $T$ gets sufficiently large so that the UAV flying time could be practically ignored, the UAV can stay stationary above each of the users to maintain best channels and maximize the throughput. On the other hand, a larger $T$ also incurs a larger access delay for users since each user may need to wait for a longer time to communicate with the UAV from one period to another.
Therefore, the period $T$ needs to be properly chosen in practice to strike a balance between the user throughput and access delay, i.e., there exists a fundamental throughput-access delay trade-off \cite{lyu2016cyclical} in UAV-enabled communications.

Without loss of generality, we consider a 3D Cartesian coordinate system where the horizontal  coordinate of the ground user $i$ is denoted by ${\mathbf{w}}_{i}=[x_i,y_i]^T \in \mathbb{R}^{2\times 1}$, $i\in \mathcal{K}$. The UAV is assumed to fly at a fixed altitude $H$ above ground and its time-varying horizonal coordinate over time is denoted by $\mathbf{q}(t)=[x(t), y(t)]^T\in \mathbb{R}^{2\times 1}$. In practice,  the UAV trajectory needs to satisfy the following two constraints:
\begin{align}
\q(0) &= \q(T), \label{eq1}\\
||\dot{\q}(t)|| &\leq V_{\max},  0\leq t \leq T, \label{eq2}
\end{align}
where (\ref{eq1}) imposes the constraint that the UAV needs to return to its initial location by the end of each period $T$ such that users can be served periodically,
 and (\ref{eq2}) corresponds to the maximum UAV speed constraint, with $\dot{\mathbf{q}}(t)$ denoting the derivative of $\mathbf{q}(t)$ with respect to $t$ and $V_{\max}$ denoting the maximum UAV speed in meter/second (m/s).

For ease of exposition, we assume that each  period $T$ is  discretized into $N$ equal-time slots, indexed by $n=1,...,N$.  The elemental slot length $\delta_t = \frac{T}{N}$ is chosen to be sufficiently small such that the UAV's location is considered as approximately unchanged within each time slot even at the maximum speed $V_{\max}$. As such, the UAV trajectory  $\mathbf{q}(t)=[x(t), y(t)]^T$ over $T$ can be approximated by  the $N$ two-dimensional sequences $\mathbf{q}[n]=[x[n], y[n]]^T$, $n=1,\cdots, N$. As a result, the trajectory constraints (\ref{eq1}) and (\ref{eq2}) can be equivalently written as
\begin{align}
\q[1] &= \q[N],\\
||\mathbf{q}[n+1]-\mathbf{q}[n]||^2 &\leq S_{\max}^2,  n=1,...,N-1,
\end{align}
where $S_{\max} \triangleq V_{\max}\delta_t$ is the maximum horizonal distance that the UAV can travel in a time slot.
Assuming that all users' locations are known, the distance from the UAV to user $i$ in time slot $n$  can be expressed as
 \begin{align}
 d_{i}[n] =  \sqrt{H^2 +||\q[n]-\w_i||^2}, \forall\, n.
 \end{align}
  For simplicity, we assume that the communication links from the UAV to the ground users are dominated by the LoS links where the channel quality depends only on the UAV-user distance. Furthermore, the Doppler effect caused by the mobility of the UAV  is assumed to be well compensated at the user receivers.  Thus, the channel power gain from the UAV to user $i$ during slot $n$ follows the free-space path loss model, which can be  expressed as
   \begin{align}
h_i[n]& = \rho_0d^{-2}_i[n] =\frac{\rho_0}{ H^2 +||\q[n]-\w_i||^2},   \forall\, n,
 \end{align}
 where  $\rho_0$ denotes the channel power gain at the reference distance $d_0=1$ m.

  Define a binary variable  $\alpha_i[n]$, which indicates that user $i$ is served by the UAV in time slot $n$ if $\alpha_i[n] = 1$; otherwise, $\alpha_i[n] = 0$. With TDMA, at most one user is scheduled for communication with the UAV  in each time slot, which yields the following constraints
  \begin{align}
  \sum_{i=1}^{K}\alpha_{i}[n]& \leq 1, \forall\, n,    \\
  \alpha_{i}[n]&\in\{0, 1\}, \forall\, i, n.
  \end{align}
Denote the transmission power of the UAV as $P$, which is assumed to be constant over time. If user $i$ is scheduled for communication in time slot $n$, the maximum  achievable rate in bits/second/Hz (bps/Hz) can be expressed as
  \begin{align}
 R_i[n] &= \log_2\left( 1 +\frac{Ph_i[n]}{\sigma^2} \right),  \nonumber\\
 & = \log_2\left( 1 +\frac{\gamma_0}{H^2+||\mathbf{q}[n]-\mathbf{w}_i||^2} \right),
 \end{align}
 where $\sigma^2$ is the additive white Gaussian noise (AWGN) power at the receiver, which is assumed to be identical for all ground users and $\gamma_0 \triangleq \frac{P\rho_0}{\sigma^2}$ denotes the reference received signal-to-noise ratio (SNR) at $d_0= 1$ m.
Thus, the achievable average rate of user $i$ over $N$ time slots is given by
   \begin{align}
R_i&= \frac{1}{N}\sum_{n=1}^{N}\alpha_i[n] R_i[n]\nonumber\\
&=\frac{1}{N}\sum_{n=1}^{N}\alpha_i[n] \log_2\left( 1 +\frac{\gamma_0}{H^2+||\mathbf{q}[n]-\mathbf{w}_i||^2} \right).
 \end{align}

\subsection{Problem Formulation}
Let $\mathbf{A}=\{\alpha_{i}[n], \forall\, i,n\}$ and $\mathbf{Q}=\{\mathbf{q}[n], \forall\,n\}$.
 Our goal is to maximize the minimum average rate among all ground users (for fairness)  by jointly optimizing the user scheduling (i.e., $\A$) and UAV trajectory (i.e., $\Q$).  Define $\eta(\A,\Q)= \min \limits_{i\in \mathcal{K}}~ R_i$ as a function of $\A$ and $\Q$. The optimization problem is formulated as
 \begin{subequations}\label{probm6}
 \begin{align}
&\max  \limits_{\mathbf{A},\mathbf{Q}} ~~ ~\eta  \nonumber\\ %\label{probm6}
  %\sum_{n=1}^{N}\alpha_{i}[n]  \log_2\left( 1 +\frac{\gamma_0}{H^2+||\mathbf{q}[n]-\mathbf{w}_i||^2} \right)     \nonumber  \\
&~\text{s.t.}  ~ \sum_{n=1}^{N}\alpha_i[n]   \log_2\left( 1 +\frac{\gamma_0}{H^2+||\mathbf{q}[n]-\mathbf{w}_i||^2} \right)  \geq \eta, \forall\, i, \label{eq012} \\
&~~~~ ~  \sum_{i=1}^{K}\alpha_{i}[n]\leq 1, \forall\, n,  \label{eq12} \\
&~~~~ ~ \alpha_{i}[n]\in\{0, 1\}, \forall\, i, n,   \label{eq13} \\
&~~~~ ~||\mathbf{q}[n+1]-\mathbf{q}[n]||^2 \leq S_{\max}^2,  n=1,...,N-1,  \label{eq14}\\
&~~~~ ~ \mathbf{q}[1]=\mathbf{q}[N]. \label{eq15}
 \end{align}
 \end{subequations}

  Problem  (\ref{probm6}) is challenging  to solve due to the following two main reasons. First,  the optimization variables $\mathbf{A}$  for user scheduling are binary and thus (\ref{eq012})-(\ref{eq13}) involve integer constraints. Second, even with fixed user scheduling variables $\mathbf{A}$, (\ref{eq012}) is still a non-convex constraint with respect to UAV trajectory variables $\Q$.
%  he optimization variables $\{\mathbf{A}, \mathbf{Q}\}$ are coupled in the objective function where no convexity structure can be exploited even with respect only  to $\mathbf{\Q}$.
Therefore, problem  (\ref{probm6}) is a mixed-integer non-convex problem, which is difficult to be optimally solved in general.  To solve this problem, we first relax the binary variables in (\ref{eq13}) into continuous variables, which yields the following problem 
 \begin{subequations}\label{probm66}
 \begin{align}
&\max  \limits_{\mathbf{A},\mathbf{Q}} ~~~ ~\eta  \nonumber\\ %\label{probm6}
  %\sum_{n=1}^{N}\alpha_{i}[n]  \log_2\left( 1 +\frac{\gamma_0}{H^2+||\mathbf{q}[n]-\mathbf{w}_i||^2} \right)     \nonumber  \\
&~~\text{s.t.}  ~~ 0\leq\alpha_i[n]\leq 1, \forall\, i,n,   \label{eq13d} \\
&~~~~~~ ~ \text{(\ref{eq012}),  (\ref{eq12}), (\ref{eq14}), (\ref{eq15})}.   \label{eq13d}
 \end{align}
 \end{subequations}
Such a relaxation in general suggests that the objective value of problem (\ref{probm66}) serves as an upper bound for that of problem  (\ref{probm6}).
 Although relaxed, problem (\ref{probm66}) is still a non-convex optimization problem due to the non-convex constraint (\ref{eq012}). In general, there is no standard method for solving such non-convex optimization problems efficiently. In the following, we first propose an efficient iterative algorithm for problem (\ref{probm66}) which is guaranteed to converge to at least a locally optimal solution and then show how to construct the solution of problem  (\ref{probm6}) based on that of problem (\ref{probm66}).
%The proposed algorithm is mainly based on exploiting the block coordinate descent method and the successive convex optimization technique.
 \section{Proposed Solution}
 In this section, we propose an iterative algorithm for problem  (\ref{probm66}) by applying the block coordinate descent and successive convex optimization techniques \cite{hong2016unified}. Specifically, for given UAV trajectory $\mathbf{Q}$, we optimize the user scheduling $\A$ by solving a linear programming (LP). On the other hand, for any given user scheduling $\A$, the UAV trajectory $\Q$ is optimized based on the successive convex optimization technique. Then, we present the overall algorithm and analytically show its convergence. Finally, we propose a low-complexity  initialization scheme for the UAV trajectory.
% Denote $r$ as the iteration number. For given $\mathbf{Q}^r$, we optimize the user scheduling to obtain $\A^{r+1}$, and then for given $\A^{r+1}$ and $\Q^r$, we optimize the UAV trajectory to obtain $\Q^{r+1}$.
%  Then, we show the objective value of problem (\ref{probm66}) increases in each iteration and the proposed algorithm is guaranteed to converge to at least a local optimal solution of problem (\ref{probm66}). Finally, we show that the obtained local optimal solution for problem (\ref{probm66}) can be utilized to construct a local optimal solution problem (\ref{probm6}) asymptotically.
 % present the overall algorithm, analytically show its convergence, and propose a scheme for the initialization of the UAV trajectory.
 %We first investigate the sub-problems of user scheduling and trajectory optimization, respectively, with the other block of variables fixed.
\subsection{User Scheduling Optimization}
%By introducing a slack variable  $t_{\rm asc} \triangleq  \min\limits_{i\in \mathcal{K}} R_i$,
For any given UAV trajectory $\Q$, problem (\ref{probm66})  is simplified as
%  \begin{align}\label{probm330}
%&~\mathop {\text{max} }\limits_{t,\A}~~~~~ t  \\
%&~~~\text{s.t.} ~~~~~  \frac{1}{N}  \sum_{n=1}^{N}\alpha_{i}[n]  \log_2\left( 1 +\gamma_i[n] \right)  \geq t_{\rm asc},\\
%&~~~~~~~~ ~~~ \sum_{i=1}^{K}\alpha_i[n]\leq 1, \forall\, n, \\
%&~~~~~~~~ ~~~ \alpha_i[n]\in\{0, 1\}, \forall\, i, n,
% \end{align}
% where $\gamma_i[n]\triangleq \frac{\gamma_0}{H^2+||\mathbf{q}[n]-\mathbf{w}_i||^2}$.
% Note that problem (P2) is a mixed integer linear programming (LP) which is in general difficult to solve optimally. By relaxing the binary variables to real  variables, problem (P2) can be rewritten as
\begin{subequations}\label{probm333}
\begin{align}
&~\mathop {\text{max} }\limits_{\eta,\A}~~~~~~~~ \eta  \nonumber\\
&~~~\text{s.t.} ~~~~~   \frac{1}{N} \sum_{n=1}^{N}\alpha_i[n]  R_i[n]  \geq \eta, \\
&~~~~~~~~ ~~~ \sum_{i=1}^{K}\alpha_i[n]\leq 1, \forall\, n,  \\
&~~~~~~~~ ~~~ 0\leq\alpha_i[n]\leq 1, \forall\, i,n.
 \end{align}
 \end{subequations}
%where $\gamma_i[n]\triangleq \frac{\gamma_0}{H^2+||\mathbf{q}[n]-\mathbf{w}_i||^2}$.
It is evident  that problem (\ref{probm333}) is a standard LP, which can be solved by existing  optimization tools such as CVX \cite{cvx}. %With the obtained optimal solution $\A^{r+1}$, it follows that
%\begin{align}\label{increase1}
%t(\A^r, \Q^r) \leq t(\A^{r+1}, \Q^r),
%\end{align}
%which suggests that the objective function value of problem (\ref{probm66}) increases in this iteration.
\subsection{Trajectory Optimization}
%By introducing a slack variable $t_{\rm trj}\triangleq  \min \limits_{i\in \mathcal{K}} R_i$,

 For any given user scheduling $\A$,  problem (\ref{probm66}) is simplified as
  \begin{subequations} \label{probm55}
    \begin{align}
 &\mathop {\text{max} }\limits_{\eta,\Q}~~~ \eta \nonumber  \\ %
&~~\text{s.t.} ~~ \sum_{n=1}^{N}\alpha_i[n]   \log_2\left( 1 +\frac{\gamma_0}{H^2+||\mathbf{q}[n]-\mathbf{w}_i||^2} \right)  \geq \eta, \forall\, i,\label{eq152} \\
&~~~~~~ ~||\mathbf{q}[n+1]-\mathbf{q}[n]||^2\leq  S_{\max}^2,  n=1,...,N-1,\label{eq27} \\
&~~~~~~ ~\mathbf{q}[1]=\mathbf{q}[N]. \label{eq151}
 \end{align}
 \end{subequations}
 %Note that problem (\ref{probm55}) is neither a concave or quasi-concave optimization problem due to the highly non-convex constraint  (\ref{eq26}). In general, there is no efficient method to obtain the optimal solution.
 Note that (\ref{eq152}) is still a non-convex constraint with respect to $\q[n]$.
 To tackle the non-convexity of  (\ref{eq152}), the successive convex optimization technique can be applied where in each iteration,  the left-hand-side (LHS) of (\ref{eq152}) is replaced by its concave lower bound at a given local point.
 Define $\Q^r =\{\mathbf{q}^r[n], \forall\,n\}$ as the given UAV trajectory in the $r$-th iteration.  The key observation is that in constraint (\ref{eq152}),  although the LHS is not concave with respect to $\mathbf{q}[n]$, it  is convex with respect to $||\mathbf{q}[n]-\mathbf{w}_i||^2$.  Recall that any convex function is globally lower-bounded by its first-order Taylor expansion at any point  \cite{bertsekas1999nonlinear}.  Therefore,  in the $r$-th iteration we obtain the following  lower bound with given local point $\mathbf{q}^r[n]$, i.e.,
   \begin{align}\label{eq155}
{R}_i[n]&=    \log_2\left( 1 +\frac{\gamma_0}{H^2+||\mathbf{q}[n]-\mathbf{w}_i||^2} \right)  \nonumber\\
& \geq -A^r_i[n]\left(||\mathbf{q}[n]-\mathbf{w}_i||^2 -||\mathbf{q}^r[n]-\mathbf{w}_i||^2 \right) + B^r_i[n] \nonumber\\
&\triangleq {R}^{{\rm lb},r}_i[n],
\end{align}
where %$\mathbf{q}^r[n]\in \Q^r$ is  the location of the UAV in time slot $n$ of the $r$-th iteration, and
\begin{align}
%B^r_i[n]&=  \frac{A^r_i[n]}{H^2+||\mathbf{q}^r[n]-\mathbf{w}_i||^2}, \forall\, i,n. \label{eq17}\\
%C^r_i[n]&=  \frac{B^r_i[n]\log_2e}{1+A^r_i[n]}, \forall\, i,n. \\
A^r_i[n]& = \frac{\gamma_0\log_2e}{(H^2+||\mathbf{q}^r[n]-\mathbf{w}_i||^2)(H^2+||\mathbf{q}^r[n]-\mathbf{w}_i||^2+\gamma_0)}, \\
B^r_i[n]&= \log_2\left(1+\frac{\gamma_0}{H^2+||\mathbf{q}^r[n]-\mathbf{w}_i||^2}\right), \forall\, i,n.
\end{align}

For any given local point $\Q^r$, define the function $\eta^{{\rm lb}, r}(\A,\Q)= \min \limits_{i\in \mathcal{K}}~ \sum_{n=1}^{N}\alpha_i[n] {R}^{{\rm lb},r}_i[n]$.  With the lower bounds ${R}^{{\rm lb}, r}_i[n]$, $\forall\, i$,  in (\ref{eq155}) and $\Q^r$, problem (\ref{probm55}) is approximated as the following problem %i%n the $r$-th iteration
\begin{subequations} \label{probm30}
 \begin{align}
 &\mathop {\text{max} }\limits_{\eta^{{\rm lb}, r},\Q}~~\eta^{{\rm lb}, r}   \nonumber\\ %
&~~~\text{s.t.} ~~ %C4: \sum_{n=1}^{N}\alpha_i[n]\left( \frac{-B^r_i[n]}{1+A^r_i[n]}\left(||\mathbf{q}[n]-\mathbf{w}_i||^2 -||\mathbf{q}^r[n]-\mathbf{w}_i||^2 \right) + \log_2\left( 1 +A^r_i[n] \right) \right) \geq t, \forall\, i, \nonumber \\
\sum_{n=1}^{N}\alpha_i[n]  {R}^{{\rm lb},r}_i[n]\geq \eta^{{\rm lb}, r}, \forall\, i,  \label{eq31} \\
&~~~~~~~~ ||\mathbf{q}[n+1]-\mathbf{q}[n]||^2\leq  S_{\max}^2,  n=1,...,N-1,  \label{eq313} \\
&~~~~~~~~ \mathbf{q}[1]=\mathbf{q}[N]. \label{eq151}
 \end{align}
 \end{subequations}
Note that both  (\ref{eq31}) and (\ref{eq313}) are convex quadratic constraints and (\ref{eq151}) is a linear constraint. Therefore, problem (\ref{probm30}) is a convex quadratically constrained quadratic program (QCQP) which
% the feasible set here is a convex set and problem (\ref{probm30}) is thereby a convex optimization problem that
can be solved efficiently by standard convex optimization solvers such as  CVX \cite{cvx}. It is worth noting that constraint (\ref{eq31}) implies (\ref{eq152}), but the reverse does not hold in general.
% the lower bound adopted in (\ref{eq31}) suggests that the feasible set of problem (\ref{probm30}) is always a subset of that of problem (\ref{probm55}), while the reverse does not hold in general.
In this regard, the optimal objective value obtained by solving problem (\ref{probm30}) always serves as a lower bound for that of problem  (\ref{probm55}). %

 \begin{algorithm}[t]
\caption{Block coordinate descent method for problem (\ref{probm66}).}\label{Algo:succ}
\begin{algorithmic}[1]
\STATE Initialize the UAV trajectory $\Q^0$. Let $r=0$.
\REPEAT
\STATE Solve problem (\ref{probm333}) for  given $\{\Q^r\}$, and denote the optimal solution as $\{\mathbf{A}^{r+1}\}$.
\STATE Solve problem (\ref{probm30}) for given $\{ \mathbf{A}^{r+1},\Q^r\}$, and denote the optimal solution as $\{\Q^{r+1}\}$.
\STATE Update $r=r+1$.
\UNTIL{  The fractional increase of the objective value  is below a threshold $\epsilon$.}
\end{algorithmic}%\vspace{-0.1cm}
\end{algorithm}\vspace{-0.1cm}
\subsection{Overall  Algorithm and Convergence}
Based on the results in the previous two subsections, we propose an overall  iterative algorithm for problem (\ref{probm66}) by applying the block coordinate descent method. Specifically, in each iteration, the user scheduling $\A$ and UAV trajectory $\Q$ are alternately optimized, by  solving either problem (\ref{probm333}) or (\ref{probm30}) correspondingly, while keeping the other block of variables fixed.
% the relaxed problem or the approximated problem with optimality, i.e.,  (\ref{probm333}) or (\ref{probm30}),
 Furthermore, the obtained solution in each iteration is used as the input of the next iteration.
The details of the algorithm are summarized in Algorithm 1.
It is worth pointing out that in the classical block coordinate descent method, the problem in each iteration is required to be solved exactly with optimality in order to guarantee the convergence \cite{bertsekas1999nonlinear}. However, in our case, for the trajectory optimization problem (\ref{probm55}), we only solve its approximated problem (\ref{probm30}) based on the lower bound in (\ref{eq155}).  Thus, the convergence analysis for the classical  coordinate descent method cannot be directly applied.

Next, we discuss the convergence of Algorithm 1 as follows. First, in step 3 of Algorithm 1,  since the optimal solution of  (\ref{probm333}) is obtained for given $\Q^r$, we have
\begin{align}\label{increase1}
\eta(\A^r, \Q^r) \leq \eta(\A^{r+1}, \Q^r),
\end{align}
where $\eta(\A, \Q)$ is defined prior to problem (\ref{probm6}).  Second, for given $\A^{r+1}$ and $\Q^r$ in step 4 of Algorithm 1, it follows that
%This suggests that the objective function value of problem (\ref{probm66}) increases in this iteration.
\begin{align}\label{increase2}
\eta(\A^{r+1}, \Q^{r}) &\overset{(a)} = \eta^{{\rm lb}, r}(\A^{r+1}, \Q^{r}) \overset{(b)} \leq \eta^{{\rm lb}, r}(\A^{r+1}, \Q^{r+1})\nonumber \\
&\overset{(c)} \leq \eta(\A^{r+1}, \Q^{r+1}),
\end{align}
where $(a)$ holds since the first-order Taylor expansion in (\ref{eq155}) is tight at the given local point which means that problem (\ref{probm30})  at  $\Q^r$ has the same objective  value as that of problem  (\ref{probm55});   $(b)$ holds since at step 4 of Algorithm 1 with the given $\A^{r+1}$, problem (\ref{probm30}) is solved optimally with solution $\Q^{r+1}$;  $(c)$ holds due to inequality (15) where for any iteration $r$, $\eta^{{\rm lb},r}(\A,\Q)$ is always a lower bound of $\eta(\A,\Q)$ for any $\A$ and $\Q$.
%since the objective  value of problem (\ref{probm30}) is the lower bound of that of its original problem (\ref{probm55}) at $\Q^{r+1}$.
The inequality in (\ref{increase2}) suggests that although only an approximated optimization problem (\ref{probm30}) is solved for obtaining the UAV trajectory, the objective  value of problem (\ref{probm55}) is still non-decreasing after each iteration.
Based on (\ref{increase1}) and (\ref{increase2}), we obtain $\eta(\A^{r}, \Q^{r})\leq \eta(\A^{r+1}, \Q^{r+1})$, which indicates that the objective value of problem (\ref{probm66}) is non-decreasing after each iteration of Algorithm 1. Since the objective value of problem  (\ref{probm66}) is upper bounded by a finite value, the proposed Algorithm 1 is guaranteed to converge. Furthermore, since the lower bound adopted in (\ref{eq155}), i.e.,  ${R}^{{\rm lb}, r}_i[n]$, has the same gradient as its original function ${R}_i[n]$ at the given local point  $\Q^r$. Thus, the convergence to a locally optimal solution  is guaranteed for Algorithm 1 based on the recent results in \cite{hong2016unified}.
 %Now, we analyze the convergence of the proposed algorithm as in the following proposition.

%\subsection{Reconstruct Binary User Scheduling}
Note that Algorithm 1 is to solve the relaxed problem (\ref{probm66}). % with the binary user scheduling variables of the original problem  (\ref{probm6}) relaxed to continuous variables between 0 and 1.
Thus, in the solution obtained by Algorithm 1, if the user scheduling variables $\alpha_{i}[n]$ are all binary, then the relaxation is tight and the obtained solution is also a locally optimal solution of problem (\ref{probm6}). Otherwise, we divide each time slot into $\tau$ sub-slots, i.e,  $N' = \tau N$, $\tau\geq 1$. Then, the number of sub-slots assigned to user $i$ in time slot $n$ is  $N_i[n]=\tau \alpha_{i}[n]$. It is not difficult to see that as $\tau$ increases, $N_i[n]$ approaches an integer which allows a binary solution.
%we round the non-binary variables to zeros or ones to construct a binary solution for problem (\ref{probm6}). %It can be shown that the objective value gap due to such a rounding approaches to zero as long as $N$ is sufficiently large.
%%between the objective value achieved by the constructed binary solution and that of the original non-binary one approaches to zero as long as $N$ is sufficiently large.
%In fact, $N$ is a controllable parameter to model the UAV trajectory and it can be always further divided into a smaller granularity such that the gap arising from the rounding decreases. %which shows the asymptotical optimality of the proposed Algorithm 1.
For example, for a two-user system with $\alpha_1[\ell]=0.69$ and $\alpha_2[\ell]=0.31$ in time slot $\ell$, they will be rounded to  1 and 0, respectively,  if $\tau=1$. If each time slot is further divided into $10$ sub-slots, i.e., $\tau=10$, then user 1 and user 2 will be assigned 6.9 and 3.1 sub-slots, respectively. Although it still leads to a non-binary solution, the gap arising from rounding $N_1[\ell]$ and $N_2[\ell]$ decreases since the duration of the sub-slot decreases. Alternatively, if each time slot is divided into $100$ sub-slots, i.e., $\tau=100$,  user 1 and user 2 will be assigned 69 and 31 sub-slots, respectively, which permits a binary solution with zero relaxation gap.
%then we have $\alpha_1[10\ell+m]=1$, $1\leq m\leq 7 $, $\alpha_2[10\ell+p]=1$, $8\leq p\leq 10 $ and
% The physical interpretation is to let the UAV remain at its current position while performing further time sharing among the users scheduled.
%Although  Algorithm 1 is proposed for solving problem (\ref{probm66}), it is able to obtain a locally optimal solution of problem (\ref{probm30}) asymptotically.
%This can be shown by the fact that for any non-binary solution of problem  (\ref{probm333}), a binary solution can be constructed by dividing the time slot  which also achieves the same objective value as the non-binary solution, i.e, without loss of optimality.
% The physical interpretation is to let the UAV remain at its current position while performing further time sharing among the users scheduled.  As a result, all users achieve equal rates for problem (\ref{probm6}).
\vspace{-0.2cm}
\subsection{Trajectory Initialization}
%It is known that for a locally optimal algorithm, the converged solution and the ultimate system performance in general depend on the initialization of the algorithm.
 In this subsection,  we propose a low-complexity  trajectory initialization scheme for Algorithm 1 based on the simple circular trajectory. Specifically, the initial UAV trajectory is set to be a circular trajectory with the UAV speed taking a constant value $V$, with $0< V\leq V_{\max}$. The trajectory circle center and radius are denoted as $\mathbf{c}_{\rm trj}=[x_{\rm trj}, y_{\rm trj}]^T$ and $r_{\rm trj}$, respectively. Then, for any given period $T$, we have $2\pi r_{\rm trj}=VT$. %Thus,  two parameters, i.e., center $C=(x_{\rm trj}, y_{\rm trj})$ and circle radius $r_{\rm trj}$, need to be specified.\left[\frac{\sum_{i=1}^{K}x_i}{K}, \frac{\sum_{i=1}^{K}y_i}{K}\right]^T
  To balance user rates, the geometric center  is a reasonable choice for the circle center of the initial UAV trajectory, i.e.,  $\mathbf{c}_{\rm trj}=\frac{\sum_{i=1}^{K}\mathbf{w}_i }{K}$. The minimum radius of a circle with $\mathbf{c}_{\rm trj}$ as the circle center which  can cover all users is denoted by $r_{\rm u}$, which is the maximum distance between  $\mathbf{c}_{\rm trj}$ and all the users, i.e.,
%\begin{align}
$r_{\rm u} = \max\limits_{i\in \mathcal K } ||\mathbf{w}_i-\mathbf{c}_{\rm trj}||.$
%\end{align}
To balance the number of users inside and outside the UAV trajectory circle, $\frac{r_{\rm u} }{2}$ is a reasonable candidate for the circle radius.
However, due to the maximum UAV speed constraint, the resulting radius $\frac{r_{\rm u} }{2}$ may not be always achievable given a finite period $T$ if $\pi r_{\rm u} >V_{\max}T$. In this case, the maximum allowed radius is computed as
%\begin{align}
$r_{\max} = \frac{V_{\max}T}{2\pi}.$
%\end{align}
As such, the radius of the initial circular trajectory is obtained as $r_{\rm trj}=\min (r_{\max},\frac{r_{\rm u} }{2} )$. Let $\theta_n \triangleq  2\pi\frac{(n-1)}{N-1}$, $\forall\, n$, and $\Q^0=\{\q^0[n], \forall\,n\}$. Based on $\mathbf{c}_{\rm trj}$ and $r_{\rm trj}$, the initial UAV trajectory in time slot $n$  is obtained as $\q^0[n] = \left[x_{\rm trj} + r_{\rm trj}\cos\theta_n,  y_{\rm trj} + r_{\rm trj}\sin\theta_n\right]^{T}$,  $n=1,...,N$.
%\begin{align}
%
%\end{align}
%%  gemometric center (-100,  238.3333)
\vspace{-0.1cm}
 \section{Numerical Results}
\begin{figure}[!t]
\centering
\includegraphics[width=0.35\textwidth]{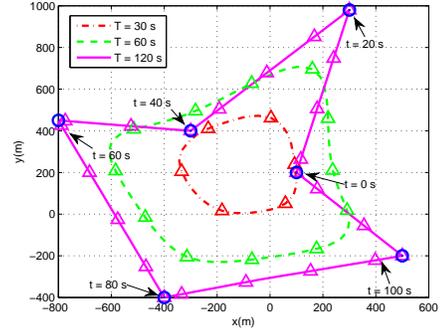}
\caption{Optimized UAV trajectories for different periods  $T$. Each trajectory is sampled every 5 s and the sampled points are marked with `$\triangle$' by using the same colors as their corresponding trajectories. The user locations are marked by blue circles `$\circ$'.} \label{single_trajectory}\vspace{-0.5cm}
\end{figure}

This section presents  numerical examples to demonstrate the effectiveness of the proposed algorithm. We consider a system with $K=6$ ground users that are randomly and uniformly distributed within a geographic area of size $1.4\times 1.4$ km$^2$. %For the purpose of illustration, we randomly generate
The following results are based on one random realization of the user locations as shown in Fig. \ref{single_trajectory}. The UAV is assumed to fly at a fixed altitude $H=100$ m. The receiver noise power is assumed to be $\sigma^2= -110$ dBm. The channel power gain at the reference distance $d_0= 1$ m is set as $\rho_0=-50$ dB. The transmit power and the maximum speed of the UAV are set as $P=0.1$ W and $V_{\max}=50$ m/s, respectively. The threshold $\epsilon$ in Algorithm 1 is set as $10^{-4}$.
%\begin{figure}[!t]
%\centering
%\includegraphics[width=0.5\textwidth]{user_location.eps}
%\caption{User distribution where the locations of user are marked with `$\diamond$'. } \label{user_location}%\vspace{-0.5cm}
%\end{figure}

%\begin{figure}[!t]
%\centering
%\includegraphics[width=0.5\textwidth]{convergence_joint2.eps}
%\caption{Convergence behaviour of the proposed Algorithm 1.} \label{single_converge}
%\end{figure}
%\subsection{Convergence of Proposed Algorithm}
%Fig. \ref{single_converge} shows the convergence behaviour of the proposed Algorithm 1. It can be observed from the figure that the max-min throughput achieved by the proposed algorithm increases quickly with the number of iterations and several tens of iterations are only needed for convergence.

   \begin{figure}[!t]
\centering
\includegraphics[width=0.35\textwidth]{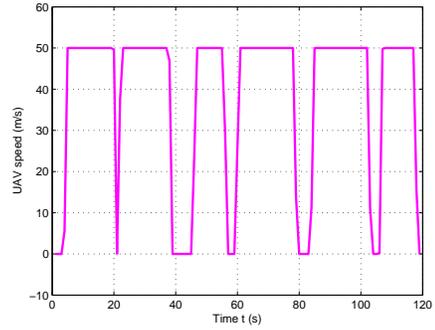}
\caption{The UAV speed versus time for $T=120$ s.} \label{speed_paper}\vspace{-0.6cm}
\end{figure}

%\begin{figure}[!t]
%\centering
%\subfigure[]{\includegraphics[width=1.5in, height=1.5in]{s1.eps}}
%\subfigure[]{\includegraphics[width=1.5in, height=1.5in]{s2.eps}}
%\subfigure[]{\includegraphics[width=3in, height=1.5in]{s3.eps}}
%\caption{UAV's speed versus different optimization horizons $T$ (s).} \label{speed_paper}\vspace{-0.5cm}
%\end{figure}

%\subsection{Convergence of Algorithm 1}
%Fig. \ref{single_converge} shows the convergence behaviour of the proposed Algorithm 1. It can be observed from the figure that the max-min throughput achieved by the proposed algorithm increases quickly with the number of iterations and several tens of iterations are only needed for convergence.
%
\vspace{-0.2cm}
\subsection{UAV Trajectory versus Cyclical Multiple Access Period  $T$}
In Fig. \ref{single_trajectory}, we illustrate the optimized trajectories obtained by the proposed Algorithm 1 under different periods $T$. It is observed that as $T$ increases, the UAV
exploits its mobility to adaptively enlarge and adjust  its trajectory to move closer to the ground users. When $T$ is sufficiently large, e.g., $T=120$ s,  the UAV is able to sequentially visit all the users and stay stationary above each user for a certain amount of time (i.e., with a zero speed), while  the UAV trajectory becomes a closed loop with segments connecting all the points right on top of the user locations. Except the time spent on traveling between the user locations, the UAV sequentially hovers above the users so as to enjoy the best communication channels. For example, for the case of $T=120$ s, it can be observed that the sampled points on the trajectory around each user have higher density than those far way from users. This means that when the UAV flies close to each user, it will reduce the speed accordingly such that more information can be transmitted over a better air-ground channel. This phenomenon can be more directly observed from Fig. \ref{speed_paper} for the case of $T=120$ s, where the UAV speed reduces  to zero when it flies right above  each user. While for $T=30$ and $60$ s, the UAV always flies at the maximum speed $V_{\max}$ in order to get as close to each user as possible for shorter LoS links within each limited period $T$.

   \begin{figure}[!t]
\centering
\includegraphics[width=0.35\textwidth]{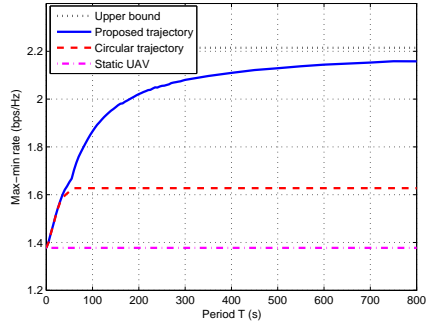}
\caption{Max-min rate  versus period $T$.} \label{single_throughput}\vspace{-0.6cm}
\end{figure}
\vspace{-0.2cm}
\subsection{Max-min Rate versus Cyclical Multiple Access Period $T$}
In Fig. \ref{single_throughput}, we compare the average max-min rate achieved by the following schemes: 1) Proposed trajectory, which is obtained by Algorithm 1; 2) Circular trajectory, which is obtained by the proposed  initialization scheme; and 3) Static UAV, where the UAV is placed at the geometric center of the users and remains static. For all the schemes, the user scheduling is optimized by Algorithm 1 with given trajectory. It is observed from Fig. \ref{single_throughput} that the max-min rate of the static UAV is independent of $T$ since without mobility, the channel links between the UAV and users are time-invariant. In contrast, for the proposed trajectory and the circular trajectory schemes, the max-min rate increases with $T$ and eventually becomes saturated when $T$ is sufficiently large. This is expected since with the UAV mobility,  a larger $T$ provides the UAV more time to fly closer to the users to be served, which thus improves the max-min rate. In addition, when $T$ and/or $V_{\max}$ is sufficiently large such that the UAV's travelling time between users is negligible, each ground user is sequentially served when the UAV is directly on top of it. In this case, the proposed algorithm achieves the performance upper bound of the max-min rate for each user, which can be obtained as
 %small portion of time is spent for the UAV's flying and most of time is spent for hovering right on top of the users. Thus, the channel link quality, although better, again becomes time-invariant and thus leads a saturated max-min data rate.
% In light of this, the saturated max-min data rate achieved by the proposed scheme is upper bounded by the one when a UAV is hovering right on top of a user. In fact, by ignoring the flying time,  the upper bound of the max-min average data rate for each user can be obtained as
\begin{align}
R^{\rm ub} = \frac{1}{K}  \log_2\left( 1 + {\frac{P\rho_0}{H^2\sigma^2}}\right)= 2.2146\  \text{bps/Hz}.
\end{align}
The asymptotic optimality of the proposed algorithm is shown as $T$ increases in Fig. \ref{single_throughput}.

By comparing the performance of the proposed trajectory with that of the circular trajectory in Fig. \ref{single_throughput},  the advantage of fully exploiting the trajectory design  is also demonstrated. Since the circular trajectory restricts the UAV to fly along a circle, the users that are not around the circle suffer from worse channels.  As a result,  more time needs to be assigned to such users, which poses the bottleneck for the achievable max-min throughput.
While for the proposed trajectory with a sufficiently large period $T$, the UAV is able to fly closer to or even stays above all users to serve them with better channels. Therefore, the max-min throughput is improved, but at the cost of longer access delay on average for the users.
%each user has the chance of enjoying short-distance LoS links, which improves the max-min rate significantly. %The performance gain becomes more evident when $T>100$ s when the time is sufficient for the UAV in the proposed scheme to hover on top of the users.
\vspace{-0.1cm}
\section{Conclusions}
In this paper, we have investigated a new UAV-enabled air-ground wireless network. The user scheduling and UAV trajectory are jointly optimized with the objective of maximizing the minimum average rate among all users.  By utilizing the block coordinate descent and successive convex optimization techniques, an efficient iterative algorithm is proposed which is guaranteed to converge to at least a locally optimal solution. Numerical results demonstrate that the UAV mobility provides the benefit of achieving better  air-ground channels and thereby improves the system throughput. Furthermore, the proposed trajectory design significantly outperforms the mobile UAV with a circular trajectory. The interesting throughput-access delay trade-off is also shown for UAV-enabled communication. Future work will investigate the general case of multiple UAVs to further improve the throughput-access delay trade-off.
\vspace{-0.1cm}
\bibliographystyle{IEEEtran}
\bibliography{IEEEabrv,mybib}

\begin{thebibliography}{10}
\providecommand{\url}[1]{#1}
\csname url@samestyle\endcsname
\providecommand{\newblock}{\relax}
\providecommand{\bibinfo}[2]{#2}
\providecommand{\BIBentrySTDinterwordspacing}{\spaceskip=0pt\relax}
\providecommand{\BIBentryALTinterwordstretchfactor}{4}
\providecommand{\BIBentryALTinterwordspacing}{\spaceskip=\fontdimen2\font plus
\BIBentryALTinterwordstretchfactor\fontdimen3\font minus
  \fontdimen4\font\relax}
\providecommand{\BIBforeignlanguage}[2]{{%
\expandafter\ifx\csname l@#1\endcsname\relax
\typeout{** WARNING: IEEEtran.bst: No hyphenation pattern has been}%
\typeout{** loaded for the language `#1'. Using the pattern for}%
\typeout{** the default language instead.}%
\else
\language=\csname l@#1\endcsname
\fi
#2}}
\providecommand{\BIBdecl}{\relax}
\BIBdecl

\bibitem{globalUAVmarket}
``Global {UAV} market.'' [Online] Available:
  \url{https://www.aiaa.org/Detail.aspx?id=33690}.

\bibitem{fb_UAV}
``Facebook takes flight.'' [Online] Available:
  \url{http://www.theverge.com/a/mark-zuckerberg-future-of-facebook/aquila-drone-internet}.

\bibitem{gl_UAV}
``Project loon.'' [Online] Available: \url{https://www.google.com/loon}.

\bibitem{zeng2016wireless}
Y.~Zeng, R.~Zhang, and T.~J. Lim, ``Wireless communications with unmanned
  aerial vehicles: Opportunities and challenges,'' \emph{{IEEE} Commun. Mag.},
  vol.~54, no.~5, pp. 36--42, May 2016.

\bibitem{mozaffari2015drone}
M.~Mozaffari, W.~Saad, M.~Bennis, and M.~Debbah, ``Drone small cells in the
  clouds: Design, deployment and performance analysis,'' in \emph{Proc. IEEE
  GLOBECOM}, 2015, pp. 1--6.

\bibitem{mozaffari2016efficient}
------, ``Efficient deployment of multiple unmanned aerial vehicles for optimal
  wireless coverage,'' \emph{{IEEE} Commun. Lett.}, vol.~20, no.~8, pp.
  1647--1650, Aug. 2016.

\bibitem{al2014optimal}
A.~Al-Hourani, S.~Kandeepan, and S.~Lardner, ``Optimal {LAP} altitude for
  maximum coverage,'' \emph{IEEE Wireless Commun. Lett.}, vol.~3, no.~6, pp.
  569--572, Dec. 2014.

\bibitem{lyu2016placement}
J.~Lyu, Y.~Zeng, R.~Zhang, and T.~J. Lim, ``Placement optimization of
  {UAV}-mounted mobile base stations,'' \emph{{IEEE} Commun. Lett.}, vol.~21,
  no.~3, pp. 604--607, Mar. 2017.

\bibitem{bor2016efficient}
R.~I. Bor-Yaliniz, A.~El-Keyi, and H.~Yanikomeroglu, ``Efficient 3-{D}
  placement of an aerial base station in next generation cellular networks,''
  in \emph{Proc. IEEE ICC}, 2016, pp. 1--5.

\bibitem{zhan2011wireless}
P.~Zhan, K.~Yu, and A.~L. Swindlehurst, ``Wireless relay communications with
  unmanned aerial vehicles: Performance and optimization,'' \emph{{IEEE} Trans.
  Aerosp. Electron. Syst.}, vol.~47, no.~3, pp. 2068--2085, Jul. 2011.

\bibitem{zeng2016throughput}
Y.~Zeng, R.~Zhang, and T.~J. Lim, ``Throughput maximization for {UAV}-enabled
  mobile relaying systems,'' \emph{{IEEE} Trans. Commun.}, vol.~64, no.~12, pp.
  4983--4996, Dec. 2016.

\bibitem{jeong2016mobile}
S.~Jeong, O.~Simeone, and J.~Kang, ``Mobile edge computing via a {UAV}-mounted
  cloudlet: Optimal bit allocation and path planning,'' \emph{arXiv preprint
  arXiv:1609.05362}, 2016.

\bibitem{zeng2016energy}
Y.~Zeng and R.~Zhang, ``Energy-efficient {UAV} communication with trajectory
  optimization,'' \emph{IEEE Trans. Wireless Commun.}, to appear, 2017.

\bibitem{lyu2016cyclical}
J.~Lyu, Y.~Zeng, and R.~Zhang, ``Cyclical multiple access in {UAV}-aided
  communications: A throughput-delay tradeoff,'' \emph{IEEE Wireless Commun.
  Lett.}, vol.~5, no.~6, pp. 600--603, Dec. 2016.

\bibitem{hong2016unified}
M.~Hong, M.~Razaviyayn, Z.-Q. Luo, and J.-S. Pang, ``A unified algorithmic
  framework for block-structured optimization involving big data: With
  applications in machine learning and signal processing,'' \emph{{IEEE} Signal
  Process. Mag.}, vol.~33, no.~1, pp. 57--77, Jan. 2016.

\bibitem{cvx}
M.~Grant and S.~Boyd, \emph{CVX: MATLAB Software for Disciplined Convex
  Programming}.\hskip 1em plus 0.5em minus 0.4em\relax Version 2.1, 2016,
  available: \url{http://cvxr.com/cvx}.

\bibitem{bertsekas1999nonlinear}
D.~P. Bertsekas, \emph{Nonlinear Programming}.\hskip 1em plus 0.5em minus
  0.4em\relax Athena Scientific, 1999.

\end{thebibliography}

\end{document}